\documentclass[epj]{webofc}
\usepackage[varg]{txfonts}   
\wocname{EPJ Web of Conferences}
\woctitle{ICNFP 2016}
\def\gsim{\mathrel{\rlap{\lower4pt\hbox{\hskip1pt$\sim$}}
 \raise1pt\hbox{$>$}}}

 \newcommand\la{\langle}
 \newcommand\ra{\rangle}
 \newcommand\beq{\begin{equation}}
 
 \newcommand\eeq{\end{equation}}
 \newcommand\beqn{\begin{eqnarray}}
 \newcommand\eeqn{\end{eqnarray}}

\def\fm{\,\mbox{fm}}
\def\GeV{\,\mbox{GeV}}
\def\TeV{\,\mbox{TeV}}
\def\lsim{\mathrel{\rlap{\lower4pt\hbox{\hskip1pt$\sim$}}
    \raise1pt\hbox{$<$}}}         
\def\gsim{\mathrel{\rlap{\lower4pt\hbox{\hskip1pt$\sim$}}
    \raise1pt\hbox{$>$}}}         

\def\Im{\,{\rm Im}\,}

\def\fm{\,\mbox{fm}}
\def\GeV{\,\mbox{GeV}}
\def\MeV{\,\mbox{MeV}}

\begin{document}
%
\selectlanguage{english}
\title{Novel scenario for production of heavy flavored mesons\\ 
       in heavy ion collisions}
%
%

\author{B.Z. Kopeliovich\inst{1}\thanks{\email{boris.kopeliovich@usm.cl}} \and
        J.   Nemchik\inst{2,3} \and
        I.K. Potashnikova\inst{1} \and
        Ivan Schmidt\inst{1}
}

\institute{Departamento de F\'{\i}sica,
Universidad T\'ecnica Federico Santa Mar\'{\i}a; and\\
Centro Cient\'ifico-Tecnol\'ogico de Valpara\'iso;
Casilla 110-V, Valpara\'iso, Chile
\and
Czech Technical University in Prague,
FNSPE, B\v rehov\'a 7,
11519 Prague, Czech Republic
\and
Institute of Experimental Physics SAS, Watsonova 47,
04001 Kosice, Slovakia
}

\abstract{
The observed strong suppression of heavy flavored hadrons 
produced with high $p_T$, is caused by final state interactions 
with the created dense medium. Vacuum radiation of high-$p_T$ 
heavy quarks ceases at a short time scale, as is confirmed 
by pQCD calculations and by LEP measurements 
of the fragmentation functions of heavy quarks. 
Production of a heavy flavored hadrons in a dense medium 
is considerably delayed due to prompt breakup 
of the hadrons by the medium. This causes 
a strong suppression of the heavy quark yield 
because of the specific shape of the fragmentation function. 
The parameter-free description is 
in a good accord with available data.}
\maketitle

%
%
%
\section{Introduction}
\label{intro}
%
%
%

In the popular scenario, explaining jet quenching, 
observed in heavy ion collisions, by induced energy loss 
in the hot medium created in the nuclear collision, 
a much weaker suppression, compared with light hadrons, 
was anticipated \cite{dk} for heavy flavors, 
caused by the dead-cone effect.

Later, however, measurements revealed similar magnitudes 
of suppression for heavy and light hadrons. 
Here we propose an alternative scenario for production 
of heavy flavored hadrons from a hot medium. 
The novel mechanism well explains data in a parameter-free way.

%
%
%
\section{Hard parton collision}
\label{sec-1}
%
%
%

\begin{figure}[h]
\hspace{0.5cm} 
 \begin{minipage}{4cm}   
\centering
\includegraphics[width=2.8cm,clip]{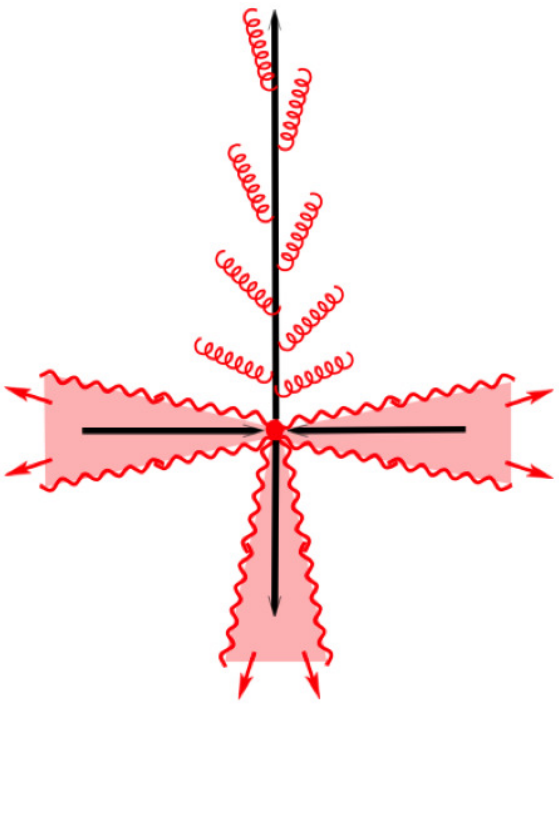}\vspace{-0.8cm}
\caption{High-$p_T$ parton scattering. The two forward-backward 
         jets are formed by the shaken-off gluon fields. 
         The high-$p_T$ partons regenerate the lost field, 
         radiating gluons.}
\label{fig:4-jet}
 \end{minipage}\hspace{0.5cm}\vspace{-0.8cm}
 \begin{minipage}{4cm}   
\centering
\includegraphics[width=4cm,clip]{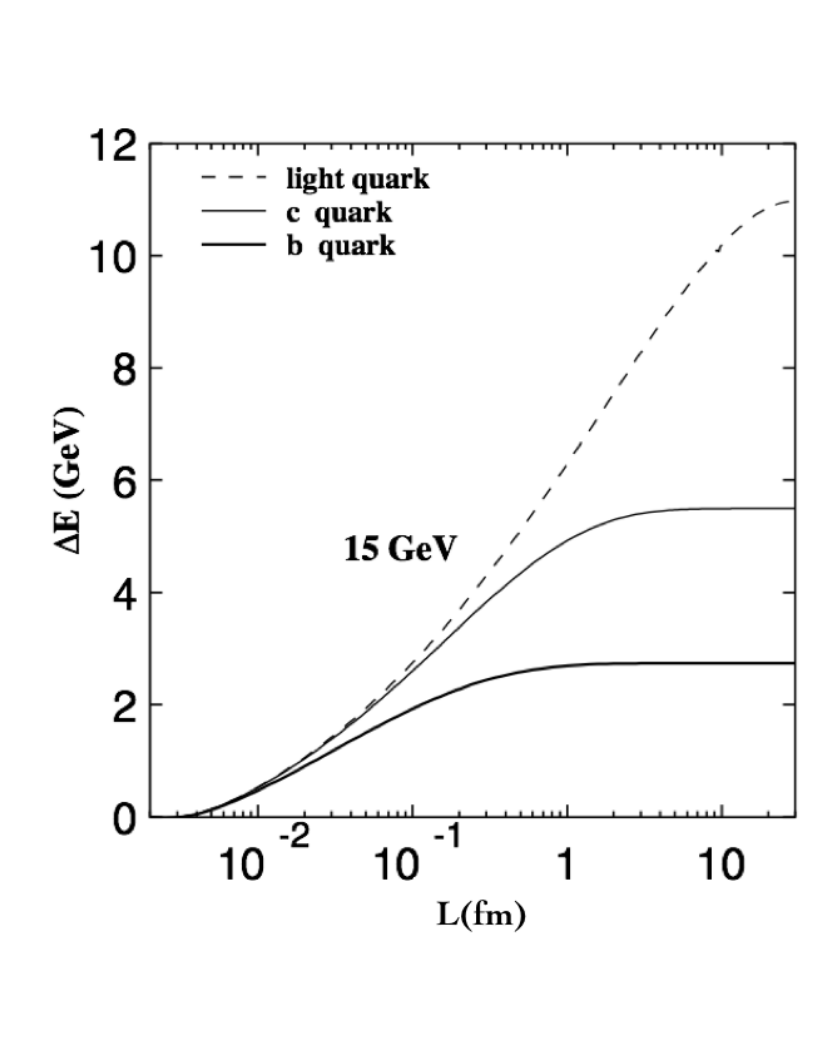}\vspace{-0.8cm}
\caption{Radiational energy loss of light, $c$ and $b$ quarks 
         having initial energy $E=\sqrt{p_T^2+m_q^2}=15\GeV$, 
         versus path length.}
\label{fig:q-c-b}
 \end{minipage}\hspace{0.5cm}
 \begin{minipage}{4cm}   
\centering
\includegraphics[width=4cm,clip]{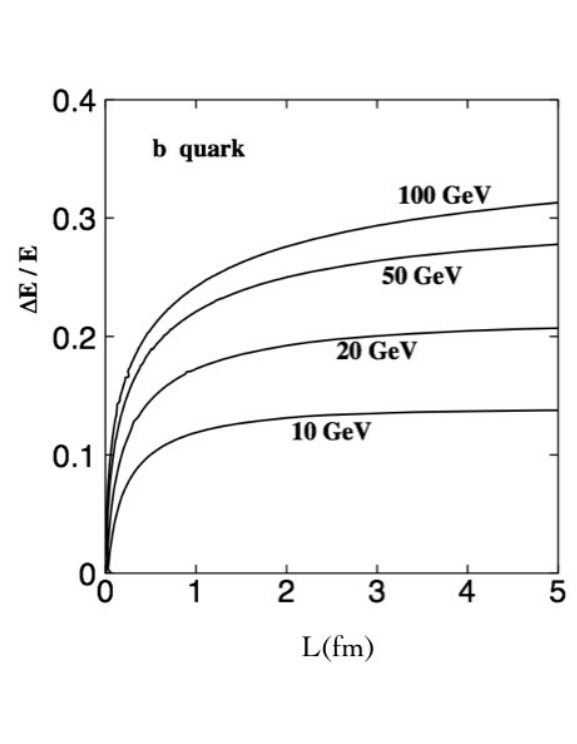}\vspace{-0.5cm}
\caption{Fractional radiational energy loss by a high-$p_T$ 
         $b$-quark, produced with different initial energies.}
\label{fig:DE_L} 
 \end{minipage}
\end{figure}

High-$p_T$ parton-parton scattering leads to 
formation of 4 cones of gluon radiation:
(i) the color field of the colliding partons is shaken off 
    in forward-backward directions;
(ii) the scattered partons carrying no field up to transverse 
    frequences $k<p_T$, are regenerating the lost components 
    of their field, radiating gluons and forming two high-$p_T$ 
    jets. This process is illustrated in figure~\ref{fig:4-jet}.
    
The coherence length/time of gluon radiation by a quark 
of mass $m_q$ and energy $E$ has the form,
\beq
l_c=\frac{2E\,x(1-x)}{k^2+x^2m_q^2}\, ,
\label{100}
\eeq
where $x$ is the fractional light-cone (LC) momentum 
of the radiated gluon. Apparently, first of all 
are radiated and regenerated gluons with small 
longitudinal and large transverse momenta.
   
The peculiar feature of these jets is closeness 
of their initial virtuality, imposed by $p_T$,
and energy $E=\sqrt{p_T^2+m_q^2}$. Therefore, 
increasing the jet energy one unavoidably 
intensifies radiation and dissipation of energy.

It is instructive to evaluate the amount of energy, 
radiated after the hard collision by the scattered 
parton over path length $L$ \cite{similar},
\beq
\Delta E_{rad}(L) =
E\int\limits_{\Lambda^2}^{p_T^2}
dk^2\int\limits_0^1 dx\,x\,
\frac{dn_g}{dx\,dk^2}
\Theta(L-l_c)\, ,
\label{210}
\eeq
where the radiation spectrum reads,
\beq
\frac{dn_g}{dx\,dk^2} =
\frac{2\alpha_s(k^2)}{3\pi\,x}\,
\frac{k^2[1+(1-x)^2]}{[k^2+x^2m_q^2]^2}\, .
\label{300}
\eeq

\begin{figure}[h]
\hspace{0.5cm} 
 \begin{minipage}{4.5cm}   
\centering
\includegraphics[width=4.5cm,clip]{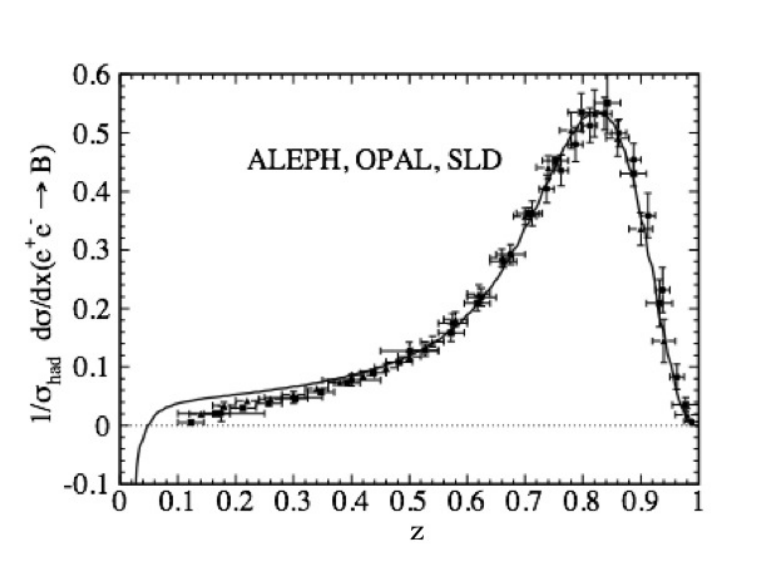}\vspace{-0.5cm}
\caption{The $b\to B$ fragmentation function, 
         from $e^+e^-$ annihilation. 
         The curve is the DGLAP fit \cite{bottom}.}
\label{fig:ff}
 \end{minipage}\hspace{0.8cm}\vspace{-0.8cm}
 \begin{minipage}{4cm}   
\centering
\includegraphics[width=4cm,clip]{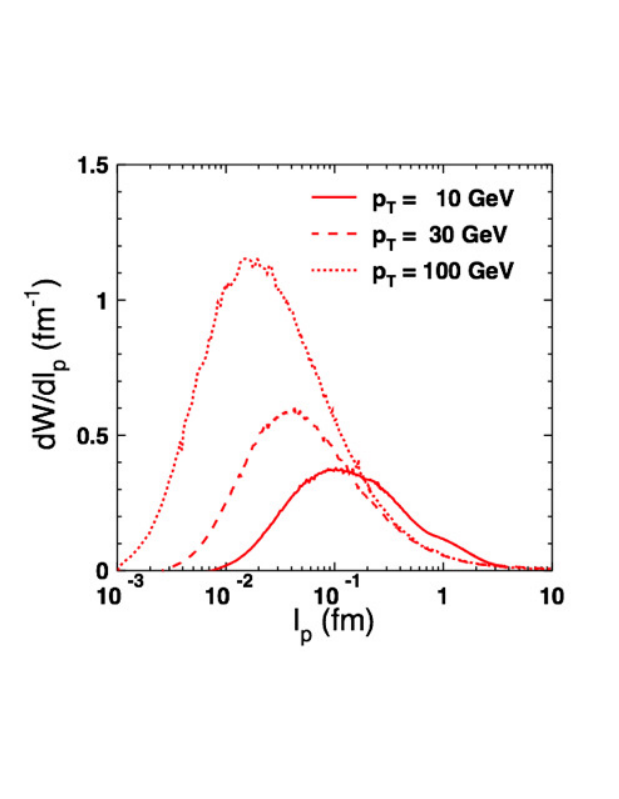}\vspace{-1.0cm}
\caption{The $l_p$-distribution of $B$-mesons produced 
         with different $p_T$ in $pp$ collisions.}
\label{fig:lp}
 \end{minipage}\hspace{0.8cm}
 \begin{minipage}{3cm}   
\centering
\includegraphics[width=2.5cm,clip]{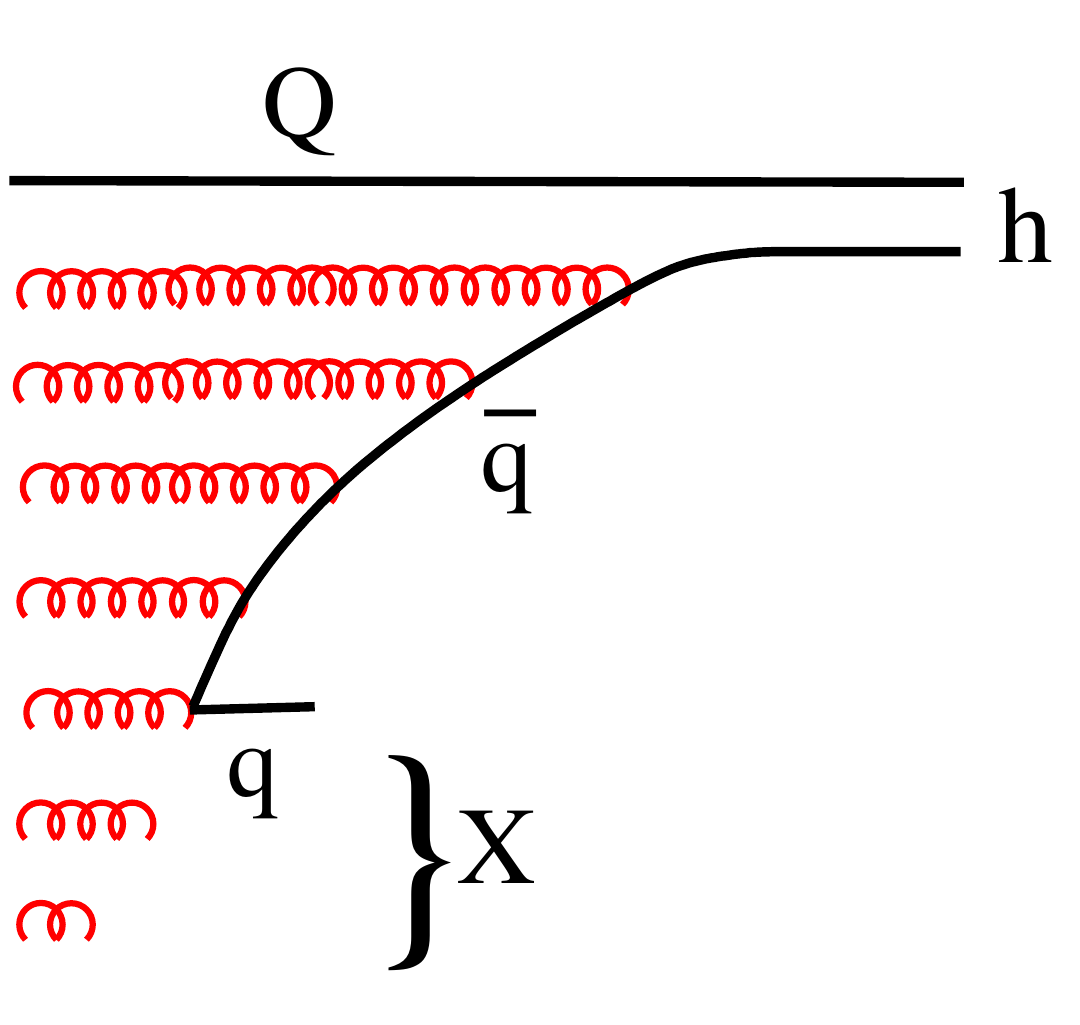}\vspace{-0.4cm}
\caption{Redistribution of the energy inside a $Q\bar q$ 
         dipole by gluon radiation by $Q$ absorbed by $\bar q$.}
\label{fig:regge-cut}
 \end{minipage}
\end{figure}

The results of calculations for absolute and fractional 
radiated energy loss are presented in figures~\ref{fig:q-c-b}
and \ref{fig:DE_L} respectively. One can see that radiation 
of heavy quarks ceases shortly. 
Only a small fraction of the initial quark energy, 
$\Delta z=\Delta E_{rad}/E$, is radiated even after 
a long time interval. 
This is quite different from the hadronization pattern 
for light quarks, which keep radiating long time and lose 
most of the initial energy (see figure~\ref{fig:q-c-b}). 
Therefore, the final $B$ or $D$ mesons carry almost 
the whole momentum of the jet. This expectation is confirmed 
by the direct measurements of the fragmentation functions 
in $e^+e^-$ annihilation \cite{bottom}. 
The example of the $b\to B$ fragmentation function  
depicted in figure~\ref{fig:ff} indeed shows that the distribution strongly 
peaks at $z\sim0.85$. A similar behavior was observed also
for the $c\to D$ fragmentation function \cite{charm}.
At the same time the fragmentation functions 
of light quarks to light mesons are well known 
to fall steadily and steeply from small $z$ towards $z=1$ \cite{kkp} 

%
%
%
\section{Production length of heavy mesons}
%
%
%

In what follows we mainly consider for concreteness 
$B$-meson production. If a $B$-meson (or a $b\bar q$ dipole) 
is produced at the length $l_p$, the momentum of $B$ equals 
to the momentum of the $b$-quark at this point. 
The fraction of the momentum carried by the light $q$ 
is very small, $x\approx m_q/m_b$, i.e. is about $5\%$.
We neglect this correction in what follows. 
Since the radiational vacuum energy loss $dE/dl$ is known, 
we can directly relate the production length distribution 
$W(l_p)$ to the $b\to B$ fragmentation function $D_{b/B}(z)$,
\beq
\frac{dW}{dl_p}=
\left.\frac{\partial \Delta p_+^b/p_+^b}{\partial l}\right|_{l=l_p}
\!D_{b/B}(z)\, ,
\label{110}
\eeq
where the production length probability distribution 
and the fragmentation function are normalized to unity,  
$\int_0^\infty dl_p\, dW/dl_p=1$ and  $\int_0^1 dz\,D_{b/B}(z)=1$ respectively;
$
z\equiv {p_+^B}/{p_+^b}=
1-{\Delta p_+^b(l_p)}/{p_+^b}
$; and 
\beq
\Delta p_+^b(l_p)=\int\limits_0^{l_p} dl\, \frac{dp_+^b(l)}{dl}\, .
\label{140}
\eeq
The rate of LC momentum loss is related to energy loss 
in accordance with $p_+^b=E+\sqrt{E^2-m_b^2}$. 
Thus, knowledge of $D_{b/B}(z)$ and $dE/dl$ gives 
a direct relation between $z$ and $l_p$. 
As far as we are able to calculate $\Delta z(L)$, 
we can extract the production length of $B$-mesons directly 
from data for $D_{b/B}(z)$.
The results for $dW/dl_p$ are plotted vs $l_p$ in figure \ref{fig:lp}.
Remarkably, the mean value of $l_p$ shrinks with rising $p_T$, 
like it happens for production of high-$p_T$ light hadrons \cite{jet-lag}.

Concluding, the fragmentation of a $b$-quark in vacuum looks 
like radiational energy loss up to a point $l=l_p$, 
where it picks up a light $\bar q$ forming a colorless 
$b\bar q$ dipole, which performs a direct transition 
to $B$ without loss the $b$-quark momentum.

%
%
%
\section{In-medium fragmentation}
%
%
%

While in vacuum production of $B$ at $l=l_p$ 
is the end of the story, in a dense medium
the $B$-meson (or $b\bar q$ dipole) can easily breakup 
interacting with the medium, and release the $b$-quark, 
which will continue hadronization and pick-up another 
light antiquark, and so on. Such recreations and breakups 
of $B$-mesons will be multiply repeated, 
until the final production of the detected $B$-meson, 
which will survive escaping from the medium.
We should understand what happens with the $b$-quark, 
while it propagates either as a constituent of a $b\bar q$ dipole, 
or is released and is losing energy to hadronization.

After the color of the $b$-quark is neutralized by $\bar q$, 
the dipole propagates without radiation and loss of energy. 
However, if the $b$-quark did not finish regeneration 
of its color field, it keeps radiating inside the dipole. 
The only difference with the preceding radiation process 
by a single quark, is that the radiation inside the dipole 
is reabsorbed by accompanying $\bar q$, as is illustrated 
in figure~\ref{fig:regge-cut}. 
Thus, the $b\bar q$ dipole does not radiate, 
its energy remains constant, however the $b$-quark energy 
is redistributed inside the dipole, 
decelerating the $b$-quark and accelerating the $\bar q$. 

As was emphasised above and demonstrated 
in figures~\ref{fig:q-c-b}, \ref{fig:DE_L},
perturbative radiation of a heavy quark ceases shortly, 
within a distance of about $1\fm$ (for a typical $p_T$ range). 
One might think, looking at figure~\ref{fig:q-c-b} 
that on longer distances the quark propagates like 
a free particle with a constant energy. 
However, this would certainly contradict confinement. 
A popular model for the non-perturbative mechanism of energy loss 
is the string model with $dE_{str}/dl = -\kappa$, 
where the string tension in vacuum is $\kappa \approx 1\GeV/fm$.

While in vacuum a heavy flavored meson is produced 
on a very short length scale, $l_p\ll 1\fm$, in a 
hot medium strong absorption pushes the 
production point to the dilute medium surface. 
Therefore non-perturbative energy loss 
becomes important for such a long-lasting  
hadronization process, which is continuing throughout 
the whole area occupied by the medium. 
However, in a deconfined hot medium no string can be formed. 
Therefore the magnitude of the string tension, 
and even its very existence, depends on 
the medium temperature. We rely on the model 
\cite{string1,string2,string3} based on the lattice simulations, 
of temperature dependence of the string tension, 
$\kappa(T)=\kappa\left(1-T/T_c\right)^{1/3}$, 
where the critical temperature is fixed at $T_c=280\MeV$.

Thus, the full rate of energy loss 
comes from both perturbative and nonperturbative mechanisms,
\beq
\frac{dE}{dl}= \frac{dE_{rad}}{dl}-\kappa(T)\, .
\label{111}
\eeq

After the $b$-quark has promptly radiated the whole spectrum 
of gluons and decreased its virtuality down to the soft scale, 
the string becomes the only source of energy loss. 
If the $b$ quark pick-up a light $\bar q$ 
they are connected by a string, which is also 
decelerating the $b$-quark and accelerating the $\bar q$ 
with a rate given by the string tension. 
Such an exchange of energy between $b$ abd $\bar q$ is similar 
to what we observed above for perturbative radiation. 
So we can conclude that the $b$-quark is constantly 
loosing energy with a rate, which does not depend on 
whether it propagates alone, or as a constituent 
of a $b\bar q$ dipole. 
The process will finalize only after the last recreation 
of a $B$-meson, which escapes from the medium without 
further breakups. Apparently, the final $B$-meson will have 
a reduced energy compared with a $B$-meson produced 
at $l=l_p$ in vacuum, where it has no possibility 
for further breakups. In other words, with the same 
starting momentum $p_+^b$ the final momentum of the $B$ 
coming out of a medium, will be smaller than in vacuum. 
This causes suppression because of steeply falling 
$p_T$ distribution of the perturbatively produced $b$-quarks, 
and due to the steep fall-off of the $b$-quark 
fragmentation function at small $z$, 
as is shown in figure~\ref{fig:ff}.
Notice that above description of fragmentation and 
time-dependent energy loss holds for charm quarks as well.

%
%
%
\section{Where does the suppression come from?}
%
%
%

The cross section of production of a $B$-meson 
with momentum $p_T$ in $pp$ collision reads,
\beq
\frac{d\sigma(pp\to BX)}{d^2p_T}=
\int d^2p_+^b\,
\frac{d\sigma(pp\to QX)}{d^2p_+^b}\,
{1\over z} D_{b/B}(z)\, ,
\label{160}
\eeq
where $p_+^b=p_T^b+\sqrt{(p_T^b)^2+m_b^2}$ 
is the initial LC momentum of the $b$-quark;
\beq
z\equiv\frac{(p_T+\sqrt{p_T^2+M_B^2})}{p_+^b}=
1-\frac{\Delta p_+^b(l_p)}{p_+^b}\, .
\label{200}
\eeq

Similar relation holds for $AA$ collisions, 
however $l_p^{AA}$, the production length of the final, 
last created colorless $Q\bar q$ dipole, 
is longer than in $pp$ collisions, 
so LC momentum loss $\Delta p_+^b$ is larger, and $z_{AA}$ is smaller.

Besides, the $B$-meson production cross section 
acquires a suppression factor $S(l_p^{AA})$, which is the survival 
probability of the $Q\bar q$ dipole, created at the point $l_p^{AA}$, 
to escape the medium without being broken-up, 
and to develop the hadronic wave function.
Thus in $AA$ collision Eq.~(\ref{160}) is modified as,
\beq
\frac{d\sigma(AA\to BX)}{d^2p_T}=
\int d^2p_+^b\,
\frac{d\sigma(pp\to QX)}{d^2p_+^b}\,
{1\over z_{AA}} D_{b/B}(z_{AA})\,S(l_p^{AA})\, ,
\label{220}
\eeq
with $z_{AA}=1-\Delta p_+^b(l_p^{AA})/p_+^b$.

The last factor $S(l_p^{AA})$ in (\ref{220}) is an important player, 
making the hadronization processes in $pp$ and $AA$ different. 
Indeed, if this factor were unity, $S=1$, then there would be no reason 
to delay production point to a longer distance 
$l_p^{AA}$ 
compared with hadronization in vacuum, equation~(\ref{200}). 
However, absorption terminates the colorless $b\bar q$ dipoles 
produced "too early", so it pushes the production point 
to the diluted surface of the hot medium,  
making the production length long, $l_p^{AA}\gg l_p$, and $z_{AA}\ll z$, 
resulting in a strong suppression of the fragmentation function 
$D_{b/B}(z_{AA})$ according to figure~\ref{fig:ff}.

Thus, the two last factors in (\ref{220}) work in opposite directions, 
causing suppression of produced $B$ or $D$ mesons:

(i) The fragmentation function $D_{b/B}(z)$, peaking at large $z$ 
    (fig.~\ref{fig:ff}), tends to reduce momentum loss 
    $\Delta p_+^b(l_p^{AA})$, selecting shorter $l_p^{AA}$.

(ii) However, a shorter $l_p^{AA}$ means a longer path length 
     for further propagation of the colorless $Q\bar q$ dipole 
     in the medium, increasing its chance to brake-up.

%
%
%
\section{Attenuation of a \boldmath$Q\bar q$ dipole in a hot medium}
%
%
%

A $Q\bar q$ pair produced perturbatively with initially 
small separation, quickly expands.
Indeed, the light quark in the $Q\bar q$-meson 
carries a tiny fraction of the momentum, $x\sim m_q/m_Q$. 
Therefore, even if the produced $b\bar q$ dipole 
has a small transverse separation, its size expands 
with a high speed, enhanced by $1/x$. The formation time 
of the $B$-meson wave function (in the medium rest frame) 
is very short,
\beq
t^B_f=\frac{\sqrt{p_T^2+m_B^2}}{2m_B\omega}\, ,
\label{400}
\eeq
where $\omega=300\MeV$ is the oscillator frequency, 
which determines the splitting of the ground state 
and the first radial excitation. 
For instance at $p_T=10\GeV$ the $B$ meson 
is formed on a distance $l_f^B=0.8\fm$. 

Thus, not a small-size $r^2\sim1/p_T^2$ dipole, 
like for light hadron production \cite{eloss-ct}, 
but a nearly formed large $Q\bar q$ dipole is 
propagating through the medium. 
It can be easily broken-up, so its mean free path 
is quite short. Indeed, the $B$-meson is nearly 
as big as a pion, $\la r_{ch}^2\ra_B = 0.378\fm^2$ 
\cite{radius}. The mean free path of such a meson 
in a hot medium is very short,
$\lambda_B\sim \left[{\hat q\, \la r_T^2\ra}\right]^{-1}$, 
where $ \la r_T^2\ra = 8\la r_{ch}^2\ra/3$. 
The so called transport coefficient $\hat q$ is the rate 
of broadening of the quark transverse momentum in the medium, 
this is why it controls the dipole absorption cross section.
For instance, at $\hat q=1\GeV^2/\fm$ (compare with \cite{eloss-ct}) 
the mean free path $\lambda_B=0.04\fm$, i.e. the $b$-quark 
propagates through the hot medium, frequently picking up 
and losing light antiquark comovers. 
Meanwhile the $b$-quark keeps losing energy with a rate, 
enhanced by medium-induced effects. Eventually the 
detected $B$-meson is formed and can survive 
in the dilute periphery of the medium.

%
\subsection{The suppression factor {\boldmath$S(l_p^{AA})$}}
%

Thus, the finally detected  $B$-meson is produced at $l=l_p^{AA}$. 
In the low energy limit the formation length (\ref{400}) is very short 
and one can use the eikonal Glauber approximation,
\beq
S(l_p^{AA})=\exp\left[-{\la r_B^2\ra\over2}\int\limits_{l_p^{AA}}^\infty dl\,
\hat q(l)\right]\, ,
\label{260}
\eeq
where $\la r_B^2\ra\equiv \la r_T^2\ra_B = 8\,\la r_{ch}^2\ra_B/3$.

In contrast, in another extreme, the high-energy limit, 
the dipole size is "frozen" by Lorentz time dilation. Then,
\beq
S(l_p^{AA})=\int d^2r\,dx\,\left|\Psi_B(r,x)\right|^2
\exp\left[- \,{r^2\over2}\int\limits_{l_p^{AA}}^\infty dl\,
\hat q(l)\right]\, ,
\label{330}
\eeq
where $\Psi_B(r,x)$ is the LC wave function of the $B$-meson.

The general description interpolating between these two limits, 
is the path-integral technique \cite{kz91},
summing  all paths of the $Q$ and $\bar q$. The result has the form,
\beq
S(l_1,l_2)\propto
\left|
\int\limits_0^1dx\int d^2r_1 d^2r_2\,
\Psi_M^\dagger(r_2,x)\,
G_{Q\bar q}(l_1,\vec r_1,x;l_2,\vec r_2,x)\,
\Psi_{in}(r_1,x)\right|^2\, ,
\label{700}
\eeq
where in the case under consideration $l_1=l_p^{AA}$, $l_2\to\infty$. 
The initial distribution amplitude $\Psi_{in}(r_1,x)$ 
is taken in the Gaussian form with mean separation 
$\la r_1^2\ra= \la r_B^2\ra$. 
The Green function $G_{Q\bar q}(l_1,\vec r_1,x;l_2,\vec r_2,x)$ 
describing propagation of the dipole between longitudinal coordinates 
$l_1,\,l_2$ with initial and final separations $\vec r_1$ and $\vec r_2$ 
respectively, satisfies the 2-dimensional LC equation,
\beq
i\frac{d}{dl_2}G_{Q\bar q}(l_1,\vec r_1,x;l_2,\vec r_2,x) =
\left[
-\frac{\Delta_{r_2}}{2\,p_T\,x\,(1-x)}
+V_{Q\bar q}(l_2,\vec r_2)\right]
G_{Q\bar q}(l_1,\vec r_1;l_2,\vec r_2)\, .
\label{800}
\eeq
The imaginary part of the LC potential is responsible for absorption,
$\Im V_{Q\bar q}(l,\vec r) = -\,\hat q(l) r^2/4$.
The real part is the phenomenological Cornell-type potential, 
adjusted to reproduce the masses and decay constants for $B$ and 
$D$ mesons \cite{yang,radius}.

%
%
%
\section{Results}
%
%
%

While the employed phenomenology allows a parameter-free description 
of data, one parameters is unavoidably present in such kind of analysis. 
This is the transport coefficient $\hat q$, which cannot be reliably 
predicted, in particular its coordinate and time dependence.
We employ here the popular model \cite{wang},
\beq
\hat q(l,\vec b,\vec\tau)=\frac{\hat
q_0\,l_0}{l}\, \frac{n_{part}(\vec b,\vec\tau)}{n_{part}(0,0)}
\,\Theta(l-l_0)\, ,
\label{900}
\eeq
where $\vec b$ is the impact parameter of nuclear collision, 
$\vec\tau$ is the impact parameter of the hard parton-parton
collision relative to the center of one of the nuclei,
$n_{part}(\vec b,\vec\tau)$ is the number of participants, and
$\hat q_0$ is the rate of broadening of a quark propagating in the
maximal medium density produced at impact parameter $\tau=0$ in
central collisions ($b=0$) at the time $t=t_0$ after
the collision. 
The time interval after the hard collision is
$t=l/v_{Q\bar q}$ where $v_{Q\bar q}$ is the speed of 
the ${Q\bar q}$ dipole.
We fixed the medium equilibration time at $t_0 =1\fm$.

In such a scheme $\hat q_0$ is the only fitted parameter, 
which, however, has been already determined from other hard processes 
in $AA$ collisions. In particular, in the analysis \cite{eloss-ct} 
of data on quenching of light high-$p_T$ hadrons it was found
$\hat q_0=1.6\GeV^2/\fm$ at $\sqrt{s}=200\GeV$ and $\hat q_0=2\GeV^2/\fm$ 
at $\sqrt{s}=2760\,GeV$.

Different sources of time-dependent medium-induced energy 
loss were added, including radiative and collisional mechanisms 
\cite{baier}. Medium-induced energy loss is much smaller than 
the vacuum one, and do not produce a dramatic effect. 
They are particularly small for heavy flavors.

\begin{figure}[h]
\hspace{0.5cm} 
 \begin{minipage}{6cm}   
\centering
\includegraphics[width=5.5cm,clip]{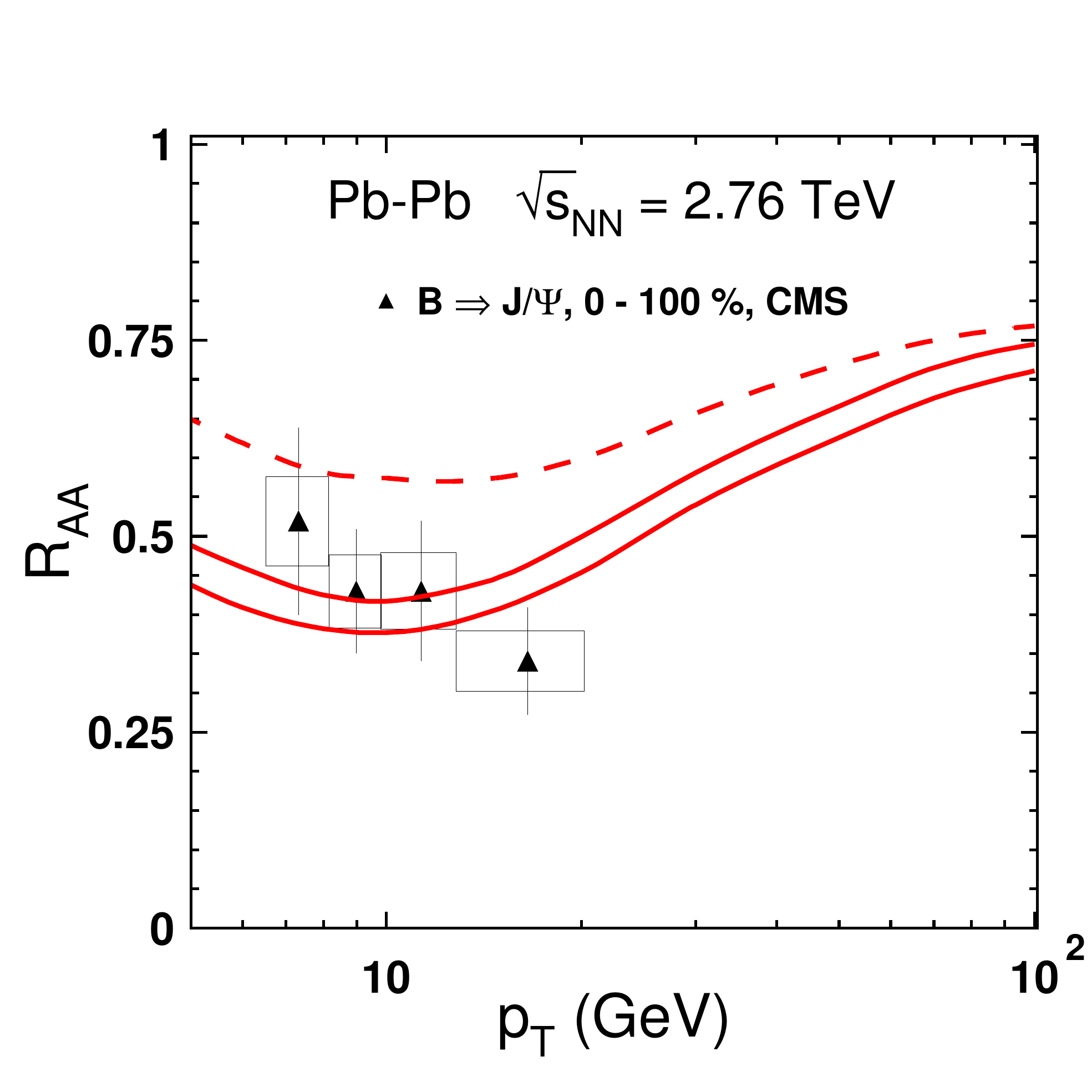}\vspace{-0.5cm}
\includegraphics[width=5.5cm,clip]{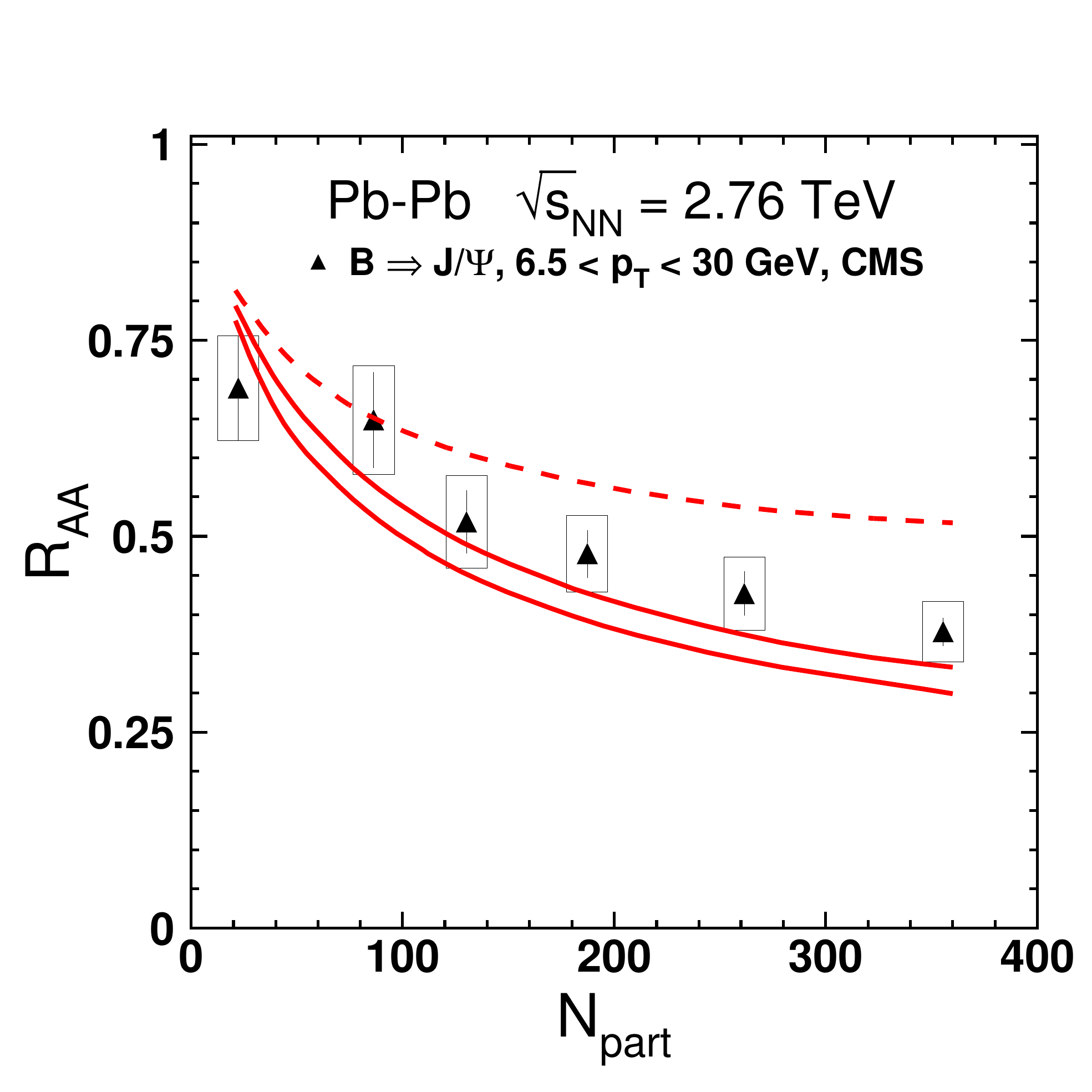}\vspace{-0.4cm}
\caption{Comparison with CMS data for indirect $J/\psi$ production 
         \cite{b-atlas} at $\sqrt{s}=2.76\TeV$. The dashed curves 
          are calculated neglecting induced energy loss. 
          The upper and bottom solid curves are calculated 
          with $\hat q_0=2.5$ and $3\GeV^2/\fm$ respectively.}
\label{fig:b-cms}
 \end{minipage}\hspace{0.5cm}
 \begin{minipage}{6cm}   
\centering
\includegraphics[width=5.5cm,clip]{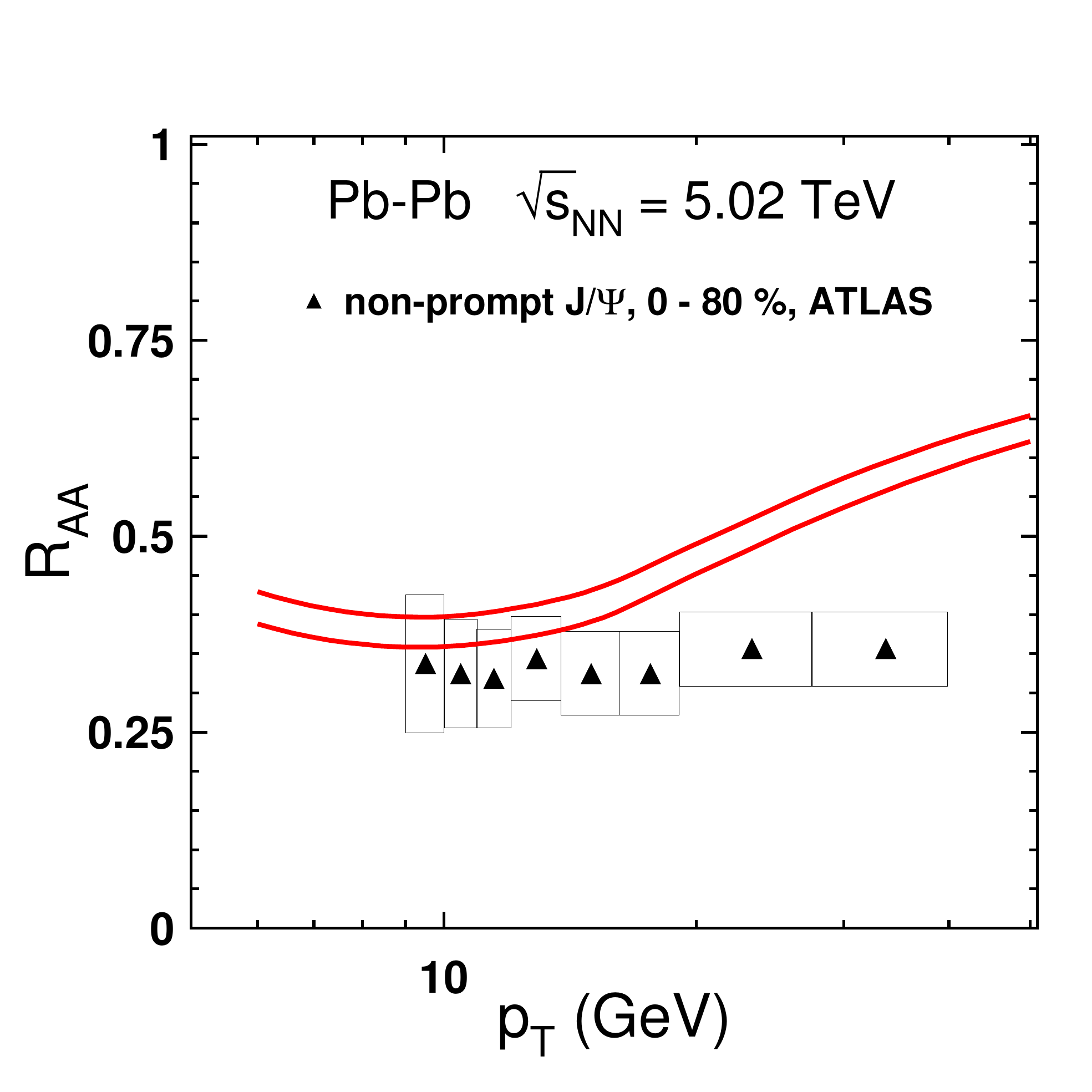}\vspace{-0.5cm}
\includegraphics[width=5.5cm,clip]{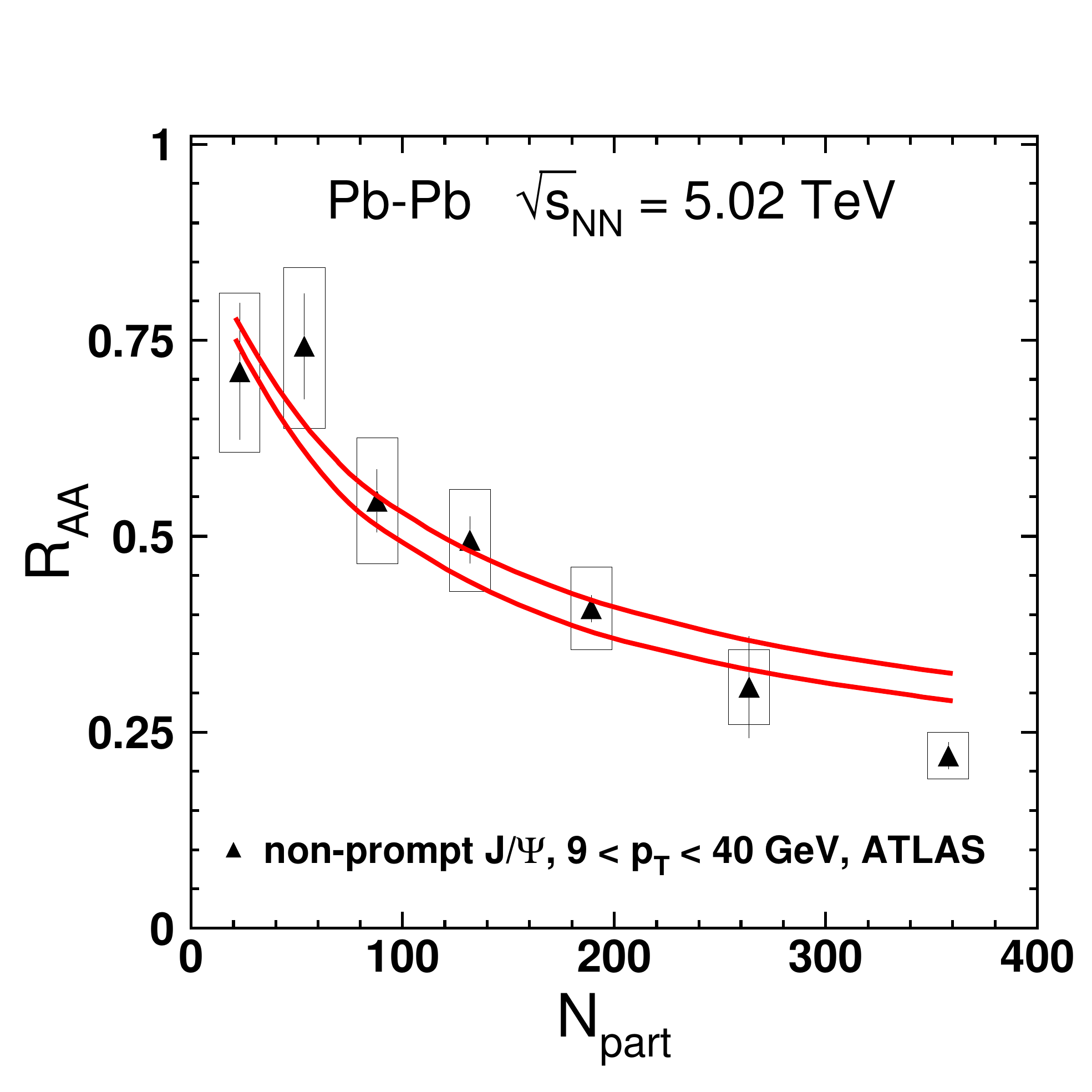}\vspace{0.4cm}
\caption{The same as in Fig.~\ref{fig:b-cms}, but at 
         $\sqrt{s}=5.02\TeV$. The upper and bottom curves 
         are calculated with $\hat q_0=2.6$ and $3.2\GeV^2/\fm$ 
         respectively. Data are from \cite{b-atlas}.}
\label{fig:b-atlas}
 \end{minipage}
\end{figure}

The results of calculations for $B$-meson production are compared 
with data on indirect production of $J/\psi$, originating 
from $B$ decays. Comparison is made vs $p_T$ and centrality.
In figure~\ref{fig:b-cms} the dashed curve is calculated 
with pure vacuum energy loss (radiative plus string) neglecting 
induced energy loss for $\hat q_0=2.5\GeV^2/\fm$. 
The upper and bottom solid curves are calculated with 
$\hat q_0=2.5$ and $3\GeV^2/\fm$ respectively, at $\sqrt{s}=2.76\TeV$ 
in comparison with data \cite{b-cms}.

Figure~\ref{fig:b-atlas} demonstrate comparison with recent data 
from ATLAS \cite{b-atlas}. The upper and bottom curves 
are calculated with $\hat q_0=2.6$ and $3.2\GeV^2/\fm$ 
respectively, at $\sqrt{s}=5.02\TeV$. 
The lack of rise of $R_{AA}$ at high $p_T$ in data looks unusual, 
compared with the typical behavior of other hadrons, so we would 
restrain of claiming a serious disagreement.

Notice that the rise of $\hat q_0$ with energy is natural, 
but its value, should not depend on the process used to measure it. 
However, different analyses of different reactions rely 
on many simplifications and model-dependent assumptions. 
Therefore, it would be naive to expect exact correspondence  
of $\hat q_0$ values extracted from data on different hard 
processes. The values found here are pretty close to those 
extracted in \cite{eloss-ct} from data on high-$p_T$ 
production of light hadrons.

\begin{figure}[h]
\hspace{0.5cm} 
 \begin{minipage}{6cm}   
\centering
\includegraphics[width=5.5cm,clip]{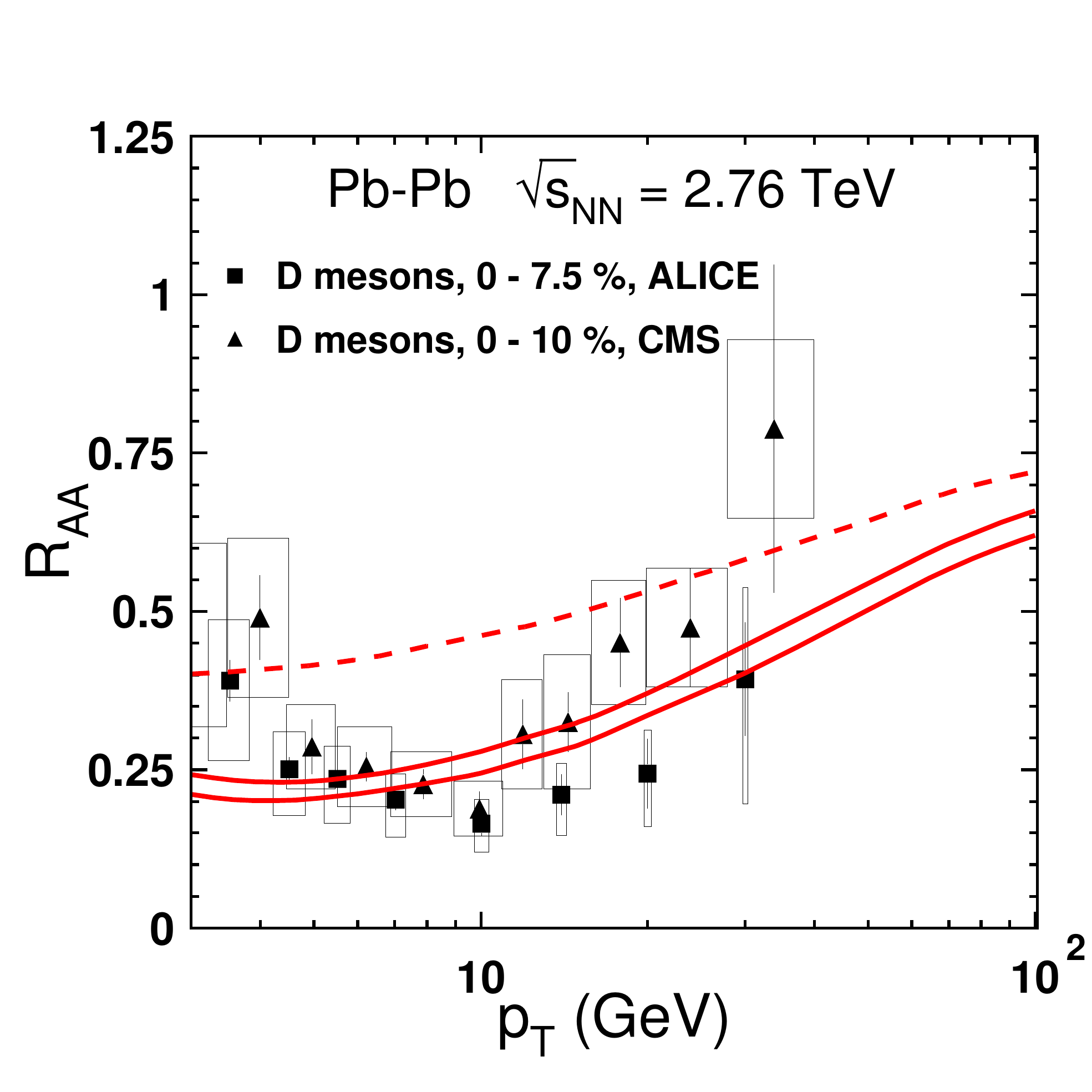}\vspace{-0.5cm}
\includegraphics[width=5.5cm,clip]{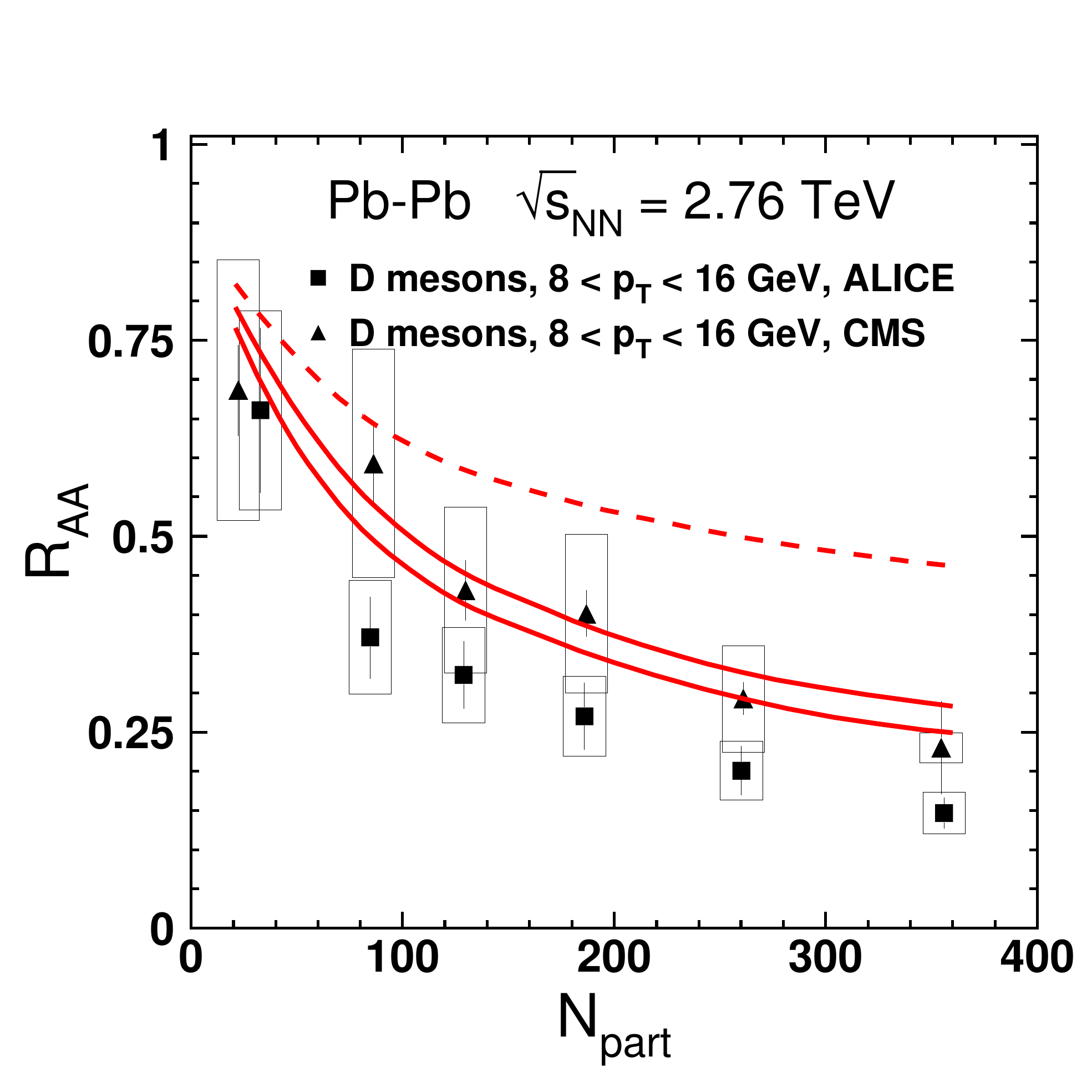}\vspace{-0.4cm}
\caption{The same as in figure~\ref{fig:b-cms}, but 
         for $D$-mesons at $\sqrt{s}=2.76\TeV$. Calculations 
         are done with the same values of $\hat q_0=2.5$ and 
         $3\GeV^2/\fm$. Data are from 
         \cite{d-alice1,d-alice2,d-alice3} and  
         \cite{d-cms1,d-cms2}.}
\label{fig:d-alice-cms}
 \end{minipage}\hspace{0.5cm}
 \begin{minipage}{6cm}   
\centering
\includegraphics[width=5.5cm,clip]{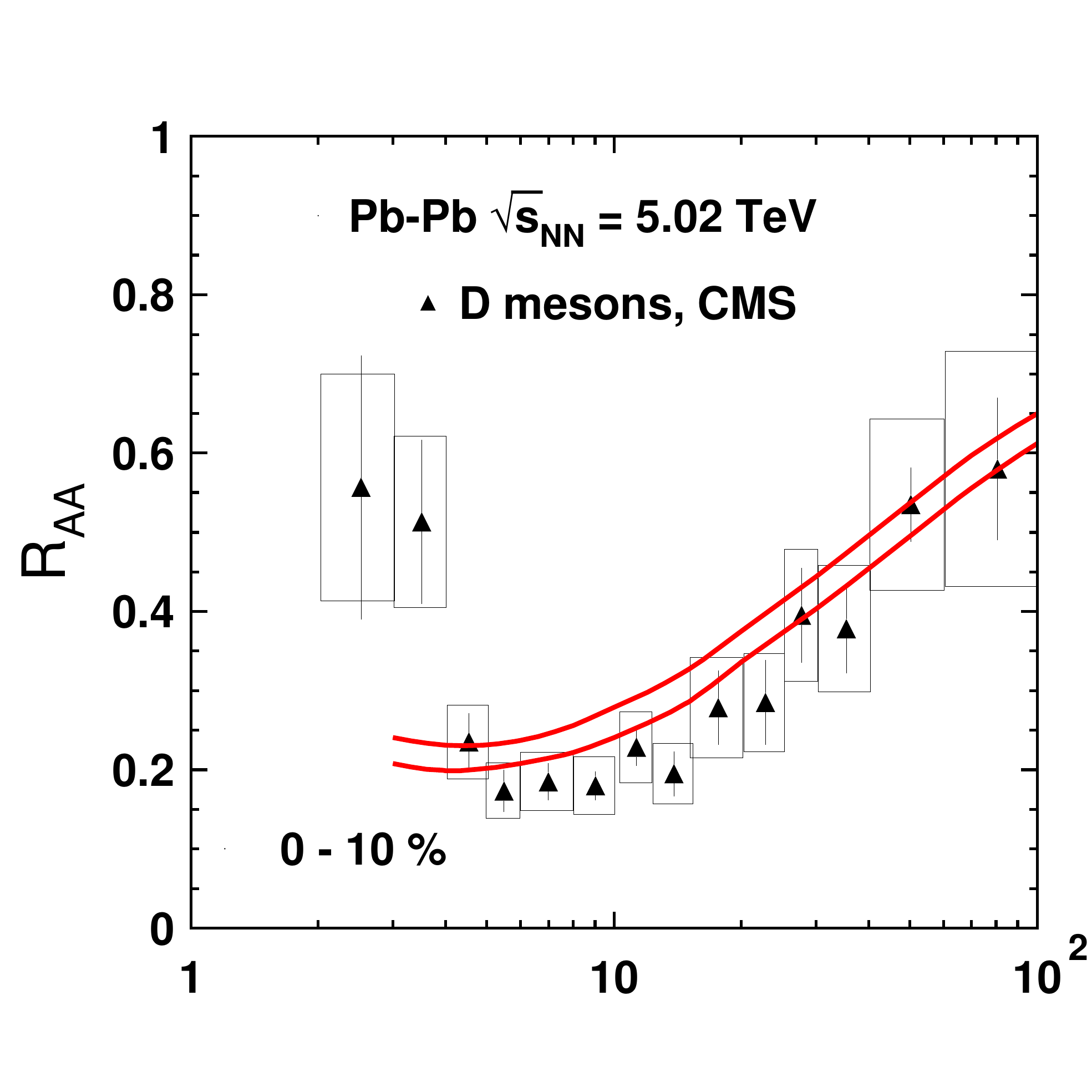}\vspace{-0.5cm}
\includegraphics[width=5.5cm,clip]{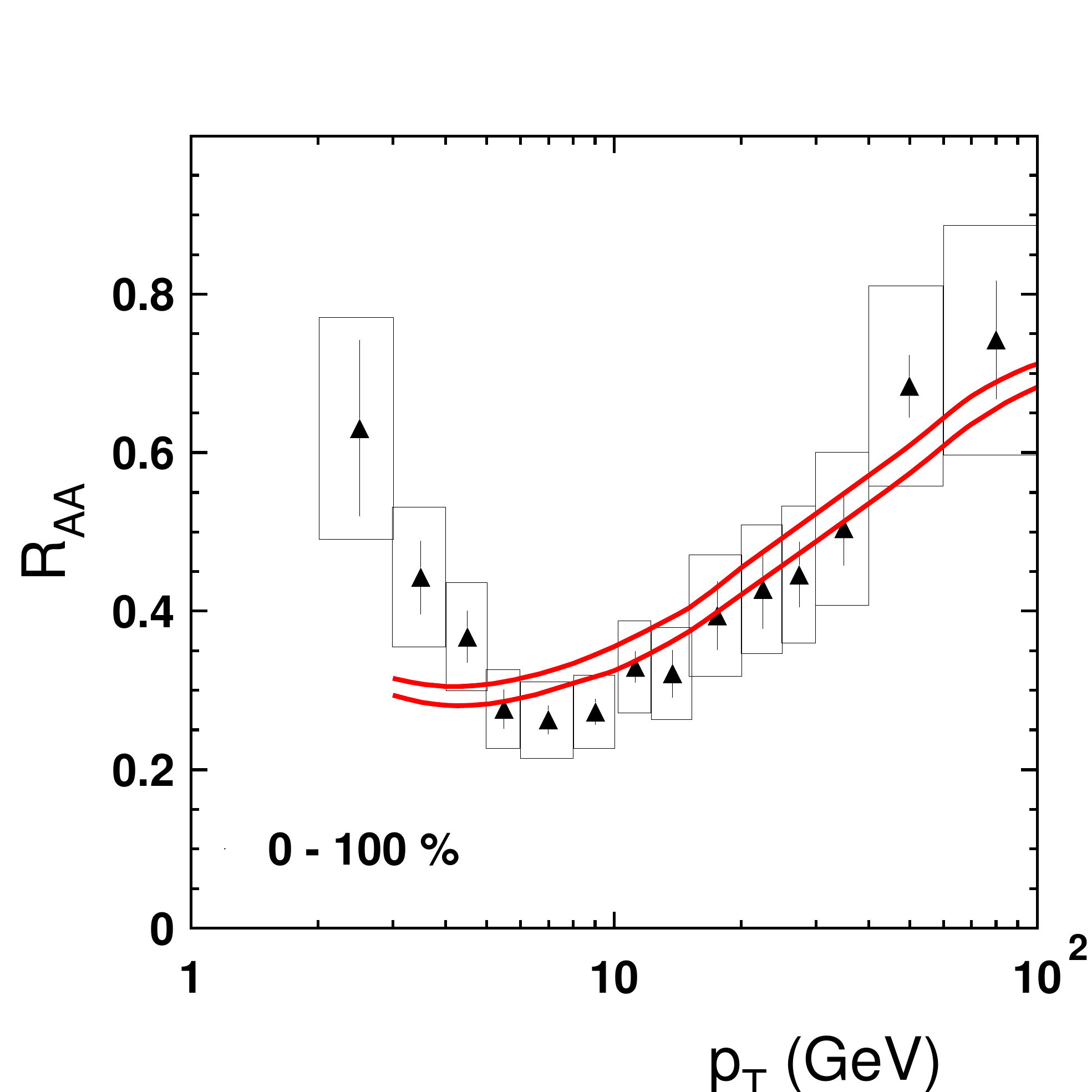}\vspace{-0.0cm}
\caption{The same as in figure~\ref{fig:d-alice-cms}, but 
         at $\sqrt{s}=5.02\TeV$ for centralities $0$-$80\%$ 
         and minimum bias events. Data from CMS measurements 
         \cite{d-cms3}.}
\label{fig:d-cms}
 \end{minipage}
\end{figure}

The approach developed here can also be applied to production 
of $D$-mesons. The results are compared with data in 
figures~\ref{fig:d-alice-cms} and \ref{fig:d-cms}
vs $p_T$ and centrality.

Notice that $c$-quarks radiate in vacuum much more energy 
than $b$-quarks, while the effects of absorption 
of $c\bar q$ and $b\bar q$ dipoles in the medium are similar. 
Therefore, $D$-mesons are suppressed in $AA$ collisions 
more than $B$-mesons.

%
%
%
\section{Summary}
%
%
%

Fragmentation of high-$p_T$ heavy quarks expose nontrivial features.

\begin{itemize}
\item
Heavy and light quarks produced in high-$p_T$ partonic collisions 
radiate differently. Heavy quarks regenerate their stripped-off 
color field much faster than light ones, and radiate a significantly 
smaller fraction of the initial energy.
\item
This peculiar feature of heavy-quark jets leads to a specific shape 
of the fragmentation functions. Differently from light flavors, 
the heavy quark fragmentation functions strongly peak 
at large fractional momentum $z$, i.e. the produced heavy-light 
meson, $B$ or $D$, carry the main fraction of the jet momentum. 
This is a clear evidence of a short production time of heavy-light 
mesons.
\item
Contrary to the propagation of a small $\bar qq$ dipole, 
which survives in the medium due to color transparency, 
a $\bar qQ$ dipole promptly expands to a large size. 
Such a big dipole has no chance to survive intact in a hot medium. 
On the other hand, a breakup of such a dipole does not suppress 
directly the production rate of $\bar qQ$ colorless dipoles, 
but increase energy loss preceding the final production 
of a heavy flavored meson. This is different from the scenario 
of high-$p_T$ production of light $\bar qq$ mesons 
\cite{eloss-ct}.
\item
Data for production of high-$p_T$ $B$ and $D$ mesons are 
explained in a parameter-free way. The extracted values 
of the transport coefficient agree with the results 
of previous analyses within unavoidable uncertainties, 
related to employed simplifications and model dependent 
assumptions, made in the calculations.
\item
We have disregarded so far the small initial state suppression 
of heavy flavors due to higher twist heavy dipole attenuation and 
leading twist shadowing \cite{kt-hf}. Inclusion of these effect will 
lead to a small decrease of the values of $\hat q_0$ extracted 
from the analysis.

\end{itemize}

\begin{acknowledgement}
This work was supported in part
by Fondecyt (Chile) grants 1130543, 1130549, 1140842, 1140377,
by Proyecto Basal FB 0821 (Chile),
and by CONICYT grant  PIA ACT1406 (Chile) .
J.N. work was partially supported
by the grant 13-20841S of the Czech Science Foundation (GA\v CR),
by the Grant M\v SMT LG15001, by the Slovak Research and Development Agency APVV-0050-11 and
by the Slovak Funding Agency, Grant 2/0020/14.

\end{acknowledgement}

\end{document}